\documentclass[12pt,a4paper]{article}
\usepackage{a4wide,axodraw}

\newcommand{\tfrac}[2]{{\textstyle\frac{#1}{#2}}}
\newcommand{\MS}{$\overline{\mathrm{MS}}$}
\newcommand{\ep}{\varepsilon}
\begin{document}

\begin{flushright}
TTP98-36
\end{flushright}
\vspace{5mm}
\begin{center}
\LARGE Decoupling of heavy-quark loops\\
in light-light and heavy-light quark currents\\[5mm]
\normalsize A.~G.~Grozin\footnote{Permanent address:
Budker Institute of Nuclear Physics,
Novosibirsk 630090, Russia.
Email: A.G.Grozin@inp.nsk.su}\\
\textit{Institute of Theoretical Particle Physics, University of Karlsruhe,}\\
\textit{D-76128 Karlsruhe, Germany}
\end{center}

\begin{abstract}
Matching of light-light quark currents in QCD with a heavy flavour
and in the low-energy effective QCD is calculated in \MS{} at two loops.
Heavy-light HQET currents are similarly considered.
\end{abstract}

Consider QCD with $n_l$ light flavours and a single heavy one having mass $M$.
Processes involving only light fields with momenta $p_i\ll M$
can be described by an effective field theory --- QCD with $n_l$ flavours
plus $1/M^n$ suppressed local operators in the Lagrangian,
which are the remnants of heavy quark loops shrinked to a point.
This low-energy effective theory is constructed to reproduce
$S$-matrix elements of the full theory expanded to some order in $p_i/M$.
This condition fixes \MS{} coupling $\alpha_s'(\mu)$ of the effective theory,
its gauge parameter $a'(\mu)$ and running light quark masses $m_i'(\mu)$
in terms of the parameters of the full theory.
Two-loop matching was considered in~\cite{BW,LRV},
and three-loop one --- in~\cite{CKS}.

Operators of the full QCD can be expanded in $1/M$ in operators
of the effective theory.
Coefficients of these expansions are fixed
by equating on-shell matrix elements.
In this Letter, I consider bilinear quark currents
$j_n(\mu)=Z_n^{-1}(\mu)j_{n0}$, where the bare currents are
$j_{n0}=\bar{q}_0\gamma^{[\mu_1}\ldots\gamma^{\mu_n]}q_0$,
and $Z_n(\mu)$ are their \MS{} renormalization constants.
In the low-energy theory, they are equal to $C_n(\mu)j_n'(\mu)$
plus power-suppressed terms (which will be omitted throughout this paper).
It is natural to match the currents at $\mu=M$,
because $C_n(M)$ contains no large logarithms.
Currents at arbitrary normalization scales can be related by
\begin{equation}
\exp\left(-\int_{\alpha_s(M)}^{\alpha_s(\mu)}
\frac{\gamma_n(\alpha)}{2\beta(\alpha)}\frac{d\alpha}{\alpha}
\right) j_n(\mu) = C_n(M)
\exp\left(-\int_{\alpha_s'(M)}^{\alpha_s'(\mu')}
\frac{\gamma_n'(\alpha)}{2\beta'(\alpha)}\frac{d\alpha}{\alpha}
\right) j_n'(\mu')\,,
\label{RG}
\end{equation}
where
\begin{equation}
\gamma_n(\alpha_s) = \frac{d\log Z_n}{d\log\mu} =
\gamma_{n0} \frac{\alpha_s}{4\pi}
+ \gamma_{n1} \left(\frac{\alpha_s}{4\pi}\right)^2 + \cdots
\label{gamma}
\end{equation}
is the current's anomalous dimension
(obtained at two loops for generic $n$ in~\cite{BG}), and
\begin{equation}
\frac{d\log\alpha_s}{d\log\mu} = -2\ep-2\beta(\alpha_s)\,,\quad
\beta(\alpha_s) = \beta_0 \frac{\alpha_s}{4\pi}
+ \beta_1 \left(\frac{\alpha_s}{4\pi}\right)^2 + \cdots\,,\quad
\beta_0 = \frac{11}{3}C_A-\frac{4}{3}T_F n_f\,,
\label{beta}
\end{equation}
$T_F=\frac{1}{2}$, $C_A=N$, $C_F=(N^2-1)/(2N)$,
$N=3$ is the number of colours; $n_f=n_l+1$;
$d=4-2\ep$ is the space-time dimension;
$\gamma_n'$, $\beta'$ are the corresponding quantities in $n_l$-flavour QCD.

On-shell matrix element of $j_n(\mu)$
is $M(\mu)=Z_q Z_n^{-1}(\mu) \Gamma_{n0}$,
where $\Gamma_{n0}$ is the bare vertex function
and $Z_q$ is the on-shell renormalization constant of the light-quark field.
It should be equal to $C_n(\mu)M_n'(\mu)$ plus power-suppressed terms,
where $M_n'(\mu)=Z_q' Z_n^{\prime-1}(\mu) \Gamma_{n0}'$.
Both matrix elements are ultraviolet-finite;
their infrared divergences coincide,
because both theories are identical in the infrared region.
Therefore (cf.~\cite{Ch,BG}),
\begin{equation}
C_n(\mu) = \frac{Z_q}{Z_q'} \frac{Z_n'(\mu)}{Z_n(\mu)}
\frac{\Gamma_{n0}}{\Gamma_{n0}'}\,.
\label{match}
\end{equation}

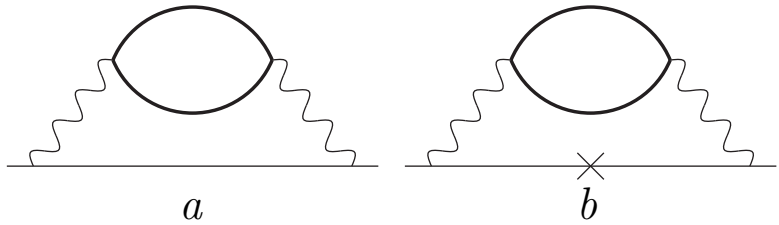
\begin{figure}[ht]
\begin{center}
\begin{picture}(290,60)
\SetWidth{1.3}
\CArc(70,27.5)(32.5,22.62,157.38)
\CArc(70,52.5)(32.5,202.62,337.38)
\SetWidth{0.5}
\Line(0,0)(140,0)
\Photon(10,0)(40,40){4}{3.5}
\Photon(100,40)(130,0){4}{3.5}
\Text(70,-20)[b]{\Large\textit{a}}
\SetWidth{1.3}
\CArc(220,27.5)(32.5,22.62,157.38)
\CArc(220,52.5)(32.5,202.62,337.38)
\SetWidth{0.5}
\Line(150,0)(290,0)
\Line(215,-5)(225,5)
\Line(215,5)(225,-5)
\Photon(160,0)(190,40){4}{3.5}
\Photon(250,40)(280,0){4}{3.5}
\Text(220,-20)[b]{\Large\textit{b}}
\end{picture}
\end{center}
\caption{Mass operator and vertex function}
\label{F}
\end{figure}

The on-shell wave-function renormalization constant is
$Z_q=\left[1-\Sigma_V(0)\right]^{-1}$,
where the bare mass operator is $\Sigma(p)=\rlap/p\Sigma_V(p^2)$.
Only the heavy-quark loop contributes (Fig.~\ref{F}\textit{a}),
and we obtain at two loops
\begin{equation}
\Sigma_V(0) = C_F T_F \frac{g_0^4 M^{-4\ep}}{(4\pi)^d} I
\frac{(d-1)(d-4)}{d}\,,
\label{Sigma}
\end{equation}
where
\begin{eqnarray}
&&i C_F g_0^2 \int \frac{d^d k}{(2\pi)^d} \frac{\Pi(k^2)}{k^4} =
C_F T_F \frac{g_0^4 M^{-4\ep}}{(4\pi)^d} I\,,
\nonumber\\
&&I = \Gamma^2(\ep) \frac{2(d-6)}{(d-2)(d-5)(d-7)} =
- \tfrac{2}{3} \Gamma^2(\ep) \left( 1 - \tfrac{2}{3}\ep + \tfrac{22}{9}\ep^2
+ \cdots \right)\,,
\label{I}
\end{eqnarray}
and $\Pi(k^2)$ is the heavy-quark contribution
to the gluon polarization operator.
This agrees with the decoupling relation~\cite{CKS}
$q=(Z_q/Z_q')^{1/2}q'$.

The bare vertex function (Fig.~\ref{F}\textit{b}) can be calculated
at any on-shell momenta of the quarks.
It is most easy to set both momenta to zero
(thus excludung power-suppressed terms in~(\ref{match})):
\begin{equation}
\Gamma_{n0} = 1 - C_F T_F \frac{g_0^4 M^{-4\ep}}{(4\pi)^d} I
\left(\frac{(d-2n)^2}{d}-1\right)\,.
\label{Gamma}
\end{equation}
The corresponding quantities in the low-energy theory
are $Z_q'=1$, $\Gamma_{n0}'=1$.
The ratio
\begin{equation}
\frac{Z_n}{Z_n'} = 1
+ \frac{1}{4} \left(\gamma_{n0}\Delta\beta_0-\Delta\gamma_{n1}\ep\right)
\left(\frac{\alpha_s}{4\pi\ep}\right)^2 =
1 + \frac{1}{9} C_F T_F \left(\frac{\alpha_s}{4\pi\ep}\right)^2
(n-1) \left[ 6(n-3) - (n-15)\ep \right]
\label{ratio}
\end{equation}
(where $\Delta\beta_0=-\frac{4}{3}T_F$ and $\Delta\gamma_{n1}$
are the contributions of a single flavour) can be obtained from~\cite{BG}.

Using
\begin{equation}
\frac{g_0^2 M^{-2\ep} \Gamma(1+\ep)}{(4\pi)^{d/2}} =
\frac{\alpha_s(M)}{4\pi} + \mathcal{O}(\alpha_s^2)\,,
\label{alpha}
\end{equation}
we finally obtain
\begin{equation}
j_n(M) = \left[ 1 + C_F T_F \left(\frac{\alpha_s}{4\pi}\right)^2
\frac{(n-1)(85n-267)}{54} \right] j_n'(M)\,.
\label{C}
\end{equation}
To all orders, $j_1=j_1'$~\cite{Ch};
this is natural, because the integral of the vector current
is the number of quarks minus antiquarks,
which is the same in both theories.
Also $m(\bar{q}q)=m'(\bar{q}q)'$ to all orders~\cite{CKS0}.
Therefore, $C_0$ can be obtained from the mass decoupling
relation~\cite{BW,CKS}.
The currents $j_3$ and $j_4$ differ from $j_1$ and $j_0$
by insertion of 't~Hooft-Veltman $\gamma_5^{\mathrm{HV}}$;
they differ from the corresponding currents with
the anticommuting $\gamma_5^{\mathrm{AC}}$ by finite renormalizations
$\bar{q}\gamma_5^{\mathrm{AC}}\gamma^\mu q=
Z_A\bar{q}\gamma_5^{\mathrm{HV}}\gamma^\mu q$,
$\bar{q}\gamma_5^{\mathrm{AC}}q=
Z_P\bar{q}\gamma_5^{\mathrm{HV}}q$,
known to three loops~\cite{L}:
\begin{eqnarray}
&&Z_A = 1 - 4 C_F \frac{\alpha_s}{4\pi}
+ C_F \left(\frac{\alpha_s}{4\pi}\right)^2
\frac{198C_F-107C_A+4T_F n_f}{9} + \cdots\,,
\nonumber\\
&&Z_P = 1 - 8 C_F \frac{\alpha_s}{4\pi}
+ C_F \left(\frac{\alpha_s}{4\pi}\right)^2
\frac{2(C_A+4T_F n_f)}{9} + \cdots
\label{ZAP}
\end{eqnarray}
Insertion of $\gamma_5^{\mathrm{AC}}$
does not change the matching coefficient.
Hence, $C_3=C_1 Z_A'/Z_A$, $C_4=C_0 Z_P'/Z_P$,
where $Z_A'$, $Z_P'$ contain $n_l$ instead of $n_f=n_l+1$.
The result for $n=2$ is, to the best of my knowledge, new.

The HQET heavy-light current
$\tilde{\jmath}(\mu)=\tilde{Z}^{-1}(\mu)\tilde{\jmath}_0$,
$\tilde{\jmath}_0=\bar{q}_0\tilde{Q}_0$ can be considered similarly
(results don't depend on its $\gamma$-matrix structure).
Instead of~(\ref{match}), we have now
\begin{equation}
\label{match2}
\tilde{C}(\mu) = \left(\frac{Z_q}{Z_q'}\right)^{1/2}
\left(\frac{\tilde{Z}_Q}{\tilde{Z}_Q'}\right)^{1/2}
\frac{\tilde{Z}'(\mu)}{\tilde{Z}(\mu)}
\frac{\tilde{\Gamma}_0}{\tilde{\Gamma}_0'}\,.
\end{equation}
On-shell renormalization constant of the static quark field is $\tilde{Z}_Q=
\left[1-\left(d\tilde{\Sigma}/d\omega\right)_{\omega=0}\right]^{-1}$,
where at two loops~\cite{BG}
\begin{equation}
\left.\frac{d\tilde{\Sigma}}{d\omega}\right|_{\omega=0} =
- C_F T_F \frac{g_0^4 M^{-4\ep}}{(4\pi)^d} (d-1) I\,.
\label{Sigma2}
\end{equation}
The bare vertex function at two loops is $\tilde{\Gamma}_0=1$~\cite{BG}.
The ratio
\begin{equation}
\frac{\tilde{Z}}{\tilde{Z}'} = 1 + C_F T_F
\left(\frac{\alpha_s}{4\pi\ep}\right)^2 \left(1-\tfrac{5}{6}\ep\right)
\label{ratio2}
\end{equation}
can be obtained from~\cite{BG0}.
Finally, we arrive at
\begin{equation}
\tilde{\jmath}(M) = \left[ 1 + \frac{89}{36} C_F T_F
\left(\frac{\alpha_s}{4\pi}\right)^2 \right] \tilde{\jmath}'(M)\,.
\label{C2}
\end{equation}

It was stated in~\cite{BG} that it is impossible to match
QCD heavy-light currents to HQET without heavy-quark loops
by equating exclusive matrix elements.
This statement is incorrect;
it was based on eq.~(2.1) of that paper,
whose right-hand side should be multiplied by $Z_q^{1/2}$
in this case.
Taking into account~(\ref{C2}), we can write the matching coefficients
for HQET without heavy-quark loops in the form~(2.20)~\cite{BG}
with $a_f'=a_f+\frac{89}{36}$.
Correspondingly, all numerical values of the coefficients
of $(\alpha_s/\pi)^2$ in Table~1~\cite{BG}
should be shifted by 0.103 in this case.

In conclusion, light-light quark currents should be adjusted
according to~(\ref{C}) when crossing a heavy-flavour threshold;
the heavy-light HQET current (say, $\bar{q}\tilde{b}$) should be adjusted
according to~(\ref{C2}) when crossing a threshold (say, of $c$ quark).

I am grateful to K.~G.~Chetyrkin for useful discussions,
and to T.~Mannel for hospitality in Karlsruhe.

\end{document}